\documentclass[aps,prd,preprint,superscriptaddress,showpacs,nofootinbib,epsf]{revtex4}
\usepackage{graphicx}% Include figure files
\usepackage{bm}% bold math

\begin{document}

\preprint{APS/123-QED}

\title{Gravitational Radiation from Rotational Core Collapse:
 Effects of Magnetic Fields and Realistic Equation of States}% Force line breaks with \\

\author{Kei Kotake}
\email[E-mail: ]{kkotake@utap.phys.s.u-tokyo.ac.jp}
\affiliation{Department of Physics, School of Science, the University of
Tokyo, 7-3-1 Hongo, Bunkyo-ku, Tokyo 113-0033, Japan}
\author{Shoichi Yamada}
\affiliation{Science \& Engineering, Waseda University, 3-4-1 Okubo, Shinjuku,
Tokyo, 169-8555, Japan}
\author{Katsuhiko Sato}
\affiliation{Department of Physics, School of Science, the University of
Tokyo, 7-3-1 Hongo, Bunkyo-ku, Tokyo 113-0033, Japan}
\affiliation{Research Center for the Early Universe, School of Science,
the University of Tokyo, 7-3-1 Hongo, Bunkyo-ku, Tokyo 113-0033, Japan}
\author{Kohsuke Sumiyoshi}
\affiliation{Numazu College of Technology, Ooka 3600, Numazu, Shizuoka
410-8501, Japan}
\author{Hiroyuki Ono}
\affiliation{Faculty of Science and Technology, Tokyo University of
Science, Yamazaki 2641, Noda, Chiba, 278-8510, Japan}
\author{Hideyuki Suzuki}
\affiliation{Faculty of Science and Technology, Tokyo University of
Science, Yamazaki 2641, Noda, Chiba, 278-8510, Japan}

\date{\today}% It is always \today, today,

\begin{abstract}
We perform a series of two-dimensional, axisymmetric, magnetohydrodynamic simulations
 of the rotational collapse of a supernova core.
 In order to calculate the waveforms of the
 gravitational wave, we derive the quadrupole
formula including the contributions from the electromagnetic
fields. Recent stellar evolution calculations imply that the magnetic
fields of the toroidal components are much stronger than those of the poloidal ones at the presupernova stage. Thus, we 
 systematically investigate the effects of the toroidal magnetic fields
 on the amplitudes and waveforms of the gravitational wave. Furthermore, we employ the two kinds of the realistic equation of
states, which are often used in the supernova simulations. 
Then, we investigate the effects of the equation of states on the
gravitational wave signals. As for the microphysics, 
we took into account electron capture and neutrino transport by the
 so-called leakage scheme. With these computations, we find that the
 peak amplitudes of the gravitational wave are lowered by an
 order of $10 \%$ for the models with the strongest toroidal magnetic fields. 
However, the peak amplitudes are
mostly within the sensitivity range of laser interferometers such as
 TAMA and the first
LIGO for a source at a distance of 10 kpc.
 Furthermore, we point out 
that the amplitudes of second peaks are still within the
 detection limit of the first LIGO for the source, although the
 characteristics of second peaks are 
 reduced by the magnetic fields.
We stress the importance of the detection, since it will give us
 information about the angular momentum 
 distribution of massive evolved stars. 
% On the other hand, it seems difficult to extract information about
% the magnetic fields only by the gravitational wave from a
% magnetorotational collapse of a supernova core.  
When we compare the gravitational waves from the two realistic equation
 of states, significant differences are not found, except that the
typical frequencies of the gravitational wave become slightly higher for the softer equation of state.

 \end{abstract}

\pacs{04.30.Db}% PACS, the Physics and Astronomy
\maketitle

\section{INTRODUCTION}
Rotation has been supposed to play an
important role in the gravitational radiation from core collapse
supernovae. The large-scale
asphericities at core bounce induced by rotation 
can convert the part of the gravitational energy 
into the form of the gravitational waves. The expected amplitude of the
gravitational waves from a supernova in the Milky Way is considered to be within the detection limits of
the long-base line laser interferometers such as [GEO600, LIGO, TAMA,
VIRGO] \cite{thorne}. The detection of the gravitational wave is important not only
for itself but also for the understanding of core collapse 
supernovae themselves, because the gravitational wave is the only window
that enables us to see directly the innermost part of an evolved massive
stars, in which the angular momentum distribution and the equation of
state are uncertain.
 
  So far there has been extensive 
work devoted to studying gravitational radiation
  in rotational core collapse \cite{mm,ys,zweg,rampp,dimmel,fry,shibata,shibaseki} (see \cite{new} for a 
review). 
More recently, Ott {\it{et al.}} \cite{ott} 
performed a large number of purely-hydrodynamic
  calculations employing a realistic equation of state (EOS) while neutrino
  transfer and microphysics are not treated. They investigated the
  effects of initial rotation rates and degree of differential rotation, on the gravitational waveforms (see also \cite{zweg,dimmel}). On the other
  hand, M\"{u}ller {\it {et al.}} \cite{muller03}
  performed a small set of the rotational core collapse simulations
  while employing the 
 elaborate neutrino transport with the detailed microphysics. Here it is noted that these  
  recent \cite{ott,muller03} studies took into account the initial 
  models based on the
  recent stellar evolution calculations \cite{heger03}, which predict      
  rather slow rotation rates at a presupernova stage than those assumed
  in the previous studies. If so, the
  peak amplitudes may not be obtained at core bounce but at  
  the later phases when the neutrino-driven convections become active 
  behind the accreting
  shock \cite{muller03}.

In addition to rotation, we study the effects of magnetic fields
on the gravitational signals in this paper. In order to estimate the
gravitational waveforms in the magnetohydrodynamic computations, we modify   
the conventional quadrupole formula including contributions from the
electromagnetic fields. 
To our knowledge, the effects have not been investigated so far.
To be rigorous, it is true that the realistic magnetorotational core collapse
simulations should require the implementation of a realistic EOS, an
adequate treatment of microphysics (weak interactions with neutrino
transfer), and relativistic treatment of gravitation in three dimensions.
However it is far beyond our scope to treat them all at once. Thus, 
we choose to employ a realistic EOS and treat the microphysics 
in a simplified manner in the Newtonian gravity assuming axial
symmetry. 
Recent stellar evolution studies show that 
the toroidal magnetic field components may be stronger than the poloidal ones
 prior to the collapse \cite{heger03,spruit}.  
We systematically change the strength of
rotation and toroidal magnetic fields in a parametric way. By so doing, we
hope to understand the effect of toroidal magnetic fields on the
waveforms and the amplitudes of the gravitational wave 
both in the weak magnetic field ($\sim 10^{12}$
G) regime and in the strong magnetic field
($\sim 10^{15}$ G) regime  at the formation of protoneutron star. The latter case may be associated with  
the formation of the so-called magnetars
\cite{duncan}, such as anomalous X-ray pulsars and soft gamma-ray repeaters
\cite{zhang,gusei}. Although the number of the magnetars is much
smaller in comparison with the canonical pulsars, the gravitational wave
from such objects should be investigated. 

Furthermore, we investigate the effects of realistic EOS's on the gravitational signals. 
Needless to say, EOS is an important microphysical ingredient for
determining the dynamics of core collapse, eventually, the gravitational
wave amplitude.  As a realistic EOS,
Lattimer-Swesty EOS \cite{Lat91} has been used in recent simulations
discussing gravitational
radiation from rotational core collapse \cite{ott,muller03}. 
It has been difficult to investigate the effects of the EOS's on the gravitational signals because available EOS's
based on different nuclear models are limited (see, however,
\cite{Bru86, Bru89a, Bru89b, Swe94}). Recently, a new complete
EOS for supernova simulations has become available \cite{shen98,shen98_2,sumi_prep}. The EOS is based on the relativistic mean
field (RMF) theory with a Thomas-Fermi approach.
 By implementing these
realistic EOS's into the magnetohydrodynamic simulations, 
we first investigate the effects of the realistic EOS's on the gravitational wave signals. 

We describe the numerical models in the next section. In the third
section, we show the numerical results. A conclusion is given in the last section.
\section{MODELS AND NUMERICAL METHODS}
\subsection{Initial Models}
Recently, Heger {\it et al}. \cite{heger03} performed the stellar evolution calculations
, in which rotation and magnetic fields are taken into account. 
Remembering caveats that their calculations are based on the 
one-dimensional models with uncertainties and not the final
 answer probably, they pointed out that the toroidal components of the magnetic
fields are stronger than the poloidal ones at the presupernova stage. 
It is because the differential rotation amplifies the toroidal components
by a convective stellar dynamo during the quasistatic stellar evolution \cite[e.g.,][]{parker,spruit}.
To our knowledge, most of the past MHD simulations for investigating the
dynamics of core collapse supernovae chosen the 
poloidal magnetic fields as the initial conditions.
This situation motivates us to take the initial conditions, 
in which the toroidal components are dominant over the poloidal
ones. Since Heger {\it et al}. \cite{heger03} calculated only a 
small set of models so far, we prefer a parametric approach to construct
the initial conditions in this paper.

We assume the following two rotation
laws. In addition, we prepare the toroidal magnetic
fields to yield the same profile as the rotation.

1. Shell-type rotation:
\begin{equation}
\Omega(r) = \Omega_0 \times \frac{{R_{0}}^2}{r^2 + R_{0}^2},
\end{equation}
\begin{equation}
B_{\phi}(r) = B_0 \times \frac{{R_{0}}^2}{r^2 + R_{0}^2},
\end{equation}
where $\Omega(r)$ is angular velocity, $B_{\phi}(r)$ is toroidal
component of the magnetic fields, $r$ is radius, and
$\Omega_0, R_{0}$ are model constants,  

2. Cylindrical rotation:
\begin{equation}
\Omega(X,Z) = \Omega_0 \times \frac{{X_{0}}^2}{X^2 + {X_{0}}^2} \cdot
\frac{Z_{0}^4}{Z^4 + Z_{0}^4},
\end{equation}
\begin{equation}
B_{\phi}(X,Z) = B_0 \times \frac{{X_{0}}^2}{X^2 + {X_{0}}^2} \cdot
\frac{Z_{0}^4}{Z^4 + Z_{0}^4},
\end{equation}
where $X$ and $Z$ denote distances from the rotational axis and the
equatorial plane, and $X_0, Z_0$ are model constants. The other parameters have the same meanings as above.

We have computed 14 models changing the combination of 
the total angular momentum, the rotation law, the degree of differential 
rotation, the total magnetic energy, and the equation of state. The model parameters are
presented in Table \ref{table:1}. The models are named after this
combination, with the first letter, S (slow), M (moderate), R (rapid) 
representing the initial $T/|W|$, the second letter S (shell-type), C
(cylindrical), denoting the rotation laws, the third letter, L
(Long), S (Short), indicating the values of $R_{0}, Z_{0}$, which
represent the degree of differential rotation, and the fourth letter,
$7 \sim 0.3$, indicating the value of $E_{\rm m}/|W|$. It is noted
that the ratio of magnetic and rotational energies to gravitational energy are
designated as $E_{\rm m}/|W|$ and $T/|W|$, respectively. We have chosen 
$\sim 10^{-3}, 0.5, 1.5  \%$ for the initial $T/|W|$ and
$\sim 10^{-7},10^{-4},10^{-2},10^{-1}, 10^{-0.3} (\sim 0.5)~ \%$ for the initial $E_{\rm m}/|W|$. 
The initial poloidal components of magnetic fields are assumed to be uniform
and parallel to the rotation axis, whose value is taken to be about four
orders of magnitude lower than the toroidal components in accordance
with the results by Heger {\it et al}. \cite{heger03}. We employ the
Lattimer-Swesty EOS in Model MSL4-LS, on the other hand, the relativistic EOS in the rest
of the models.  We have made precollapse models by taking a density, internal energy,
electron fraction distributions from the spherically symmetric $15
M_{\odot}$ model by Woosley and Weaver \cite{woos} and adding the angular
momentum and the magnetic field according to the prescription stated above.

Heger {\it et al}. \cite{heger03} pointed out that the iron core may rotate more slowly
with the magnetic fields than without. It is because the magnetic
braking reduces the angular momentum of the core during the quasistatic stellar evolution.
Model SSL7 corresponds to the magnetorotational
progenitor by them. We note that the models with the strongest magnetic
fields are probably unrealistic as suggested from the results by Heger
{\it et al}. \cite{heger03}. However we prepared these models in order to cover the wide range of the field strength and to see the effects of the magnetic fields on the gravitational wave signals clearly.

\begin{table}
\caption{The model parameters.}
\label{table:1}
\begin{center} 
\begin{tabular}{lccccccc} \hline \hline
Model & $T/|W| (\%) $  & $E_{\rm m}/|W| (\%) $   
 & Rotation Law   &
$ R_{0}$, $X_{0}$, $Z_0$ $\times 10^8$ (cm) &
$\Omega_0\,\,(\rm{s}^{-1})$   & $B_0\,\,(\rm{G})$ 
 \\ \hline
 SSL7& $3.2 \times 10^{-3}$ & $9.7 \times 10^{-8}$
          & Shell-type        &  $R_0 = 1 $  & 0.1   & $5.0 \times 10^{9}$ \\
 MSL4& $5.0 \times 10^{-1}$ & $10^{-4}$ 
          & Shell-type        &  $R_0 = 1 $  & 4.0   & $1.6 \times 10^{11}$ \\
 MSL2& $5.0 \times 10^{-1}$ & $10^{-2}$          
          & Shell-type        &  $R_0 = 1$   & 4.0   & $1.6 \times 10^{12}$ \\
 MSL1& $5.0 \times 10^{-1}$ & $10^{-1}$    
          & Shell-type        &  $R_0 =1 $   & 4.0   & $5.0 \times 10^{12}$ \\ 
 MSS4& $5.0 \times 10^{-1}$ & $10^{-4}$
          & Shell-type         &  $R_0 =0.1$  & 63.4  & $3.5 \times 10^{12}$ \\
 MSS2& $5.0 \times 10^{-1}$ & $10^{-2}$
          & Shell-type         &  $R_0 =0.1$  & 63.4  & $3.5 \times 10^{13}$ \\
 MSS1& $5.0 \times 10^{-1}$ & $10^{-1}$
          & Shell-type         &  $R_0 =0.1$  & 63.4  & $1.1 \times 10^{14}$ \\
 MCS4& $5.0 \times 10^{-1}$ & $10^{-4}$
          & Cylindrical      &  $X_0 = 0.1, Z_0 =1$   & 44.4  & $1.5 \times 10^{12}$ \\
 MCS2& $5.0 \times 10^{-1}$ & $10^{-2}$
          & Cylindrical      &  $X_0 = 0.1, Z_0 =1$   & 44.4  & $1.5 \times 10^{13}$ \\
 MCS1& $5.0 \times 10^{-1}$ & $10^{-1}$
          & Cylindrical      &  $X_0 = 0.1, Z_0 =1$   & 44.4  & $4.8 \times 10^{13}$  \\
MCS0.3 & $5.0 \times 10^{-1}$ & $10^{-0.3} \sim 0.5$
          & Cylindrical      &  $X_0 = 0.1, Z_0 =1$   & 44.4  & $1.0 \times 10^{14}$  \\
 RCS1   & 1.5               & $10^{-1}$
          & Cylindrical      &  $X_0 = 0.1, Z_0 =1$   & 76.8  & $4.8 \times 10^{13}$  
 \\
 RCS0.3 & 1.5  & $10^{-0.3} \sim 0.5$
          & Cylindrical      &  $X_0 = 0.1, Z_0 =1$   & 76.8  & $1.0 \times 10^{14}$  \\ \hline
 MSL4-LS & $5.0 \times 10^{-1}$ & $10^{-4}$ 
          & Shell-type        &  $R_0 = 1 $  & 4.0   & $1.6 \times 10^{11}$ 
\\\hline \hline
\end{tabular}
\end{center}
\end{table}

\subsection{Magnetohydrodynamics}
The numerical method for MHD computations employed in this paper is based on the ZEUS-2D code \cite{stone}.
The basic evolution equations are written as follows, 
\begin{equation}
\frac{d\rho}{dt} + \rho \nabla \cdot \mbox{\boldmath$v$} = 0,
\label{lenzoku}
\end{equation}
\begin{equation}
\rho \frac{d \mbox{\boldmath$v$}}{dt} = - \nabla P - \rho \nabla \Phi + \frac{1}{4 \pi}
(\nabla \times \mbox{\boldmath$B$})\times \mbox{\boldmath$B$},
\label{undo}
\end{equation}
\begin{equation}
\rho \frac{d \displaystyle{\Bigl(\frac{e}{\rho}\Bigr)}}{dt} = - P \nabla \cdot \mbox{\boldmath$v$},
\end{equation}
\begin{equation}
\frac{\partial \mbox{\boldmath$B$}}{\partial t} = \nabla \times ( \mbox{\boldmath$v$}\times \mbox{\boldmath$B$})
\label{ind}
\end{equation}
\begin{equation}
\Delta \Phi = 4 \pi G \rho,
\end{equation}
where $\rho, \mbox{\boldmath$v$}, e, P, \mbox{\boldmath$B$}, \Phi$ are density, velocity, internal
energy, pressure, magnetic field, gravitational potential, respectively.
We denote the Lagrangian derivative as $d/dt$ .  The ZEUS-2D is an
Eulerian code based on the finite-difference method and employs an
artificial viscosity of von Neumann and Richtmyer to capture shocks.
The time evolution of magnetic field is solved by the
induction equation, Eq. (\ref{ind}). In so doing, the code utilizes 
the so-called constrained transport (CT) method, which ensures the
divergence free ($\nabla \cdot \mbox{\boldmath$B$} = 0$) of the
numerically evolved magnetic fields at all times. Furthermore, the method of characteristics (MOC) is
implemented to propagate accurately all modes of MHD waves.
The self-gravity is managed by solving the Poisson equation with the
incomplete Cholesky decomposition conjugate gradient 
method. In all the computations, spherical coordinates are used and 
one quadrant of the meridian
section is covered with 300 ($r$) $\times$ 50 ($\theta$) mesh points.
We made several major changes to the base code to include the 
microphysics. First, we added an equation for the electron fraction to
treat electron captures and neutrino transport by the so-called leakage
scheme \cite{ep,blud,van1,van2}. The calculation of electron fraction
is done separately from the main hydrodynamic step in an
operator-splitting manner. Second, we implemented a relativistic EOS
\cite{shen98} or Lattimer-Swesty EOS \cite{Lat91}
instead of the ideal gas EOS assumed in the original code. For a more
detailed description of the methods, see Kotake {\it et al}. \cite{kotakeaniso}.

\subsection{Gravitational wave signal from the magnetized stellar cores}
Applying the methods \cite{mm,ys,yamasawa}, we derive the quadrupole
formula in order to compute the
gravitational waveforms from the magnetized stellar cores. The dimensionless gravitational wave amplitude
$h_{\mu \nu} \equiv g_{\mu \nu} - \eta_{\mu \nu}$ can be calculated by
the quadrupole formula as follows:
\begin{equation}
h_{ij}^{\rm TT}(R) = \frac{2G}{c^4}\frac{1}{R}\frac{d^2}{dt^2}I_{ij}^{TT}
\Biggl(t-\frac{R}{c}\Biggr),
\label{quadru}
\end{equation}
where $i,j$ run from 1 to 3, $t$ is the time, $R$ is the distance
from the source to the observer, the superscript ``TT'' means to take
the transverse traceless part, and $I_{ij}$ is the reduced quadrupole
defined as 
\begin{equation}
I_{ij} = \int \rho_{*} (x,t) \Biggl(x_i x_j - \frac{1}{3}x^2 \delta_{ij}
\Biggr)\,d^3 x, 
\label{quad}
\end{equation}
where $\rho_{*}$ represents the total energy density including the
contribution from the magnetic field,
\begin{equation}
\rho_{*} = \rho + \frac{B^2}{8 \pi c^2}.
\label{modrho}
\end{equation}
Expanding the right hand side of Eq. (\ref{quadru}) in terms of the pure-spin tensor
harmonics assuming axial symmetry shows that there is one nonvanishing
quadrupole term, namely $A_{20}^{E2}$. Then one derives for the component of $h^{\rm TT}$ the following formula,
\begin{equation}
h_{\theta \theta}^{\rm{TT}} = \frac{1}{8}\Biggl(\frac{15}{\pi}\Biggl)^{1/2} {\sin}^2 \alpha \,\,\frac{A_{20}^{\rm{E} 2}} {R},
\label{htt}
\end{equation}
where $\alpha$ is the angle between the symmetry axis and the line of
the sight of the observer. Here $A_{20}^{\rm{E} 2}$ is defined by
the second time derivative of the mass quadrupole formula:
\begin{equation}
A_{20}^{\rm{E} 2} = \frac{d^2}{dt^2} M_{20}^{\rm{E} 2}, 
\end{equation}
where the mass quadrupole formula is given as  
\begin{equation}
M_{20}^{\rm{E} 2} = \frac{G}{c^4}\frac{32 \pi^{3/2}}{\sqrt{15}}\int_{0}^1 d\mu
\int_{0}^{\infty} dr~\rho_{*} \Bigl(\frac{3}{2} \mu^2 - \frac{1}{2}\Bigr) r^4, 
\end{equation}
where $\mu = \cos \theta$.
By a straightforward, however tedious, calculation to replace the time
derivatives by the spatial derivatives applying the continuity equation, 
Eq. (\ref{lenzoku}), the equation of motion, 
Eq. (\ref{undo}), and the
induction equation, Eq. (\ref{ind}) noting the divergence-free constraint
($\nabla \cdot \mbox{\boldmath$B$} = 0$),  $A_{20}^{\rm{E} 2}$ can be
transformed into the following form,
\begin{equation}
 A_{20}^{\rm{E} 2} \equiv  {A_{20}^{\rm{E} 2}}_{\rm, quad} 
%+  {A_{20}^{\rm{E} 2}}_{\rm ,AV}
+  {A_{20}^{\rm{E} 2}}_{\rm ,Mag},
\end{equation}
where ${A_{20}^{\rm{E} 2}}_{\rm, quad}$ is the contribution from the matter:
\begin{eqnarray}
 {A_{20}^{\rm{E} 2}}_{\rm,quad} &=& \frac{G}{c^4} \frac{32 \pi^{3/2}}{\sqrt{15 }} 
\Biggl(\int_{0}^{1}d\mu 
\int_{0}^{\infty}  r^2  \,dr \,\rho [ {v_r}^2 ( 3 \mu^2 -1) + {v_{\theta}}^2 ( 2 - 3 \mu^2)
 - {v_{\phi}}^{2} - 6 v_{r} v_{\theta} \,\mu \sqrt{1-\mu^2} \nonumber\\
 & & - r \partial_{r} \Phi (3 \mu^2 -1) + 3 \partial_{\theta} \Phi \,\mu
\sqrt{1-\mu^2}]
 + \nonumber\\
& & \int_{0}^{1} d\mu
\int_{0}^{\infty}  r^3  \,dr [q_{r}(3\mu^2 -1) - 3~q_{\theta}~\mu 
\sqrt{1 - \mu^2}]
\Biggr), 
\label{quad}
\end{eqnarray}
%${A_{20}^{\rm{E} 2}}_{\rm, AV}$ is the contribution from the viscosity:
% \begin{eqnarray}
%{A_{20}^{\rm{E} 2}}_{\rm ,AV}= \frac{G}{c^4} \frac{32 \pi^{3/2}}{\sqrt{15 }} \%int_{0}^{1} d\mu
%\int_{0}^{\infty}  r^3  \,dr [q_{r}(3\mu^2 -1) - 3~q_{\theta}~\mu 
%\sqrt{1 - \mu^2}],
%\end{eqnarray}
${A_{20}^{\rm{E} 2}}_{\rm Mag} \equiv {A_{20}^{\rm{E} 2}}_{j \times B} +
{A_{20}^{\rm{E} 2}}_{\rho_{\rm m}}$ is the contribution from the magnetic field:
\begin{eqnarray}
{A_{20}^{\rm{E} 2}}_{j\times B} &=& 
\frac{G}{c^4} \frac{32 \pi^{3/2}}{\sqrt{15}} 
\int_{0}^{1} d\mu \int_{0}^{\infty} r^3 \,dr  \Bigl[(3 \mu^2 - 1)~ 
\frac{1}{c}~(\mbox{\boldmath$j$}\times \mbox{\boldmath$B$})_{r} - 
\nonumber \\
& & 3 \mu \sqrt{1 - \mu^2}~\frac{1}{c}~(\mbox{\boldmath$j$} \times \mbox{\boldmath$B$})_{\theta}\Bigl], 
\label{jB}
\end{eqnarray}
\begin{eqnarray}
{A_{20}^{\rm{E} 2}}_{\rho_{\rm m}} &=&\frac{G}{c^4} \frac{32 \pi^{3/2}}{\sqrt{15}}  \int_{0}^{1}d\mu \int_{0}^{\infty} \,dr ~\frac{1}{8 \pi c}\frac{d}{dt}
\Biggl[
\frac{\partial}{\partial \theta}[B_{r} r^3 (3~\mu^2 - 1)]E_{\phi} - 
\frac{\partial}{\partial r}[B_{\theta} r^3(3\mu^2 -1)]rE_{\phi} + 
\nonumber\\
&+& 
\frac{\partial}{\partial r}[B_{\phi}   r^3(3 \mu^2 - 1)]rE_{\theta}
- \frac{1}{\sin \theta}\frac{\partial}{\partial \theta}[B_{\phi}\sin \theta
r^3 (3 \mu^2 - 1)]E_{r}\Biggr]. 
\label{rhom}
\end{eqnarray}
${A_{20}^{\rm{E} 2}}_{j \times B},
{A_{20}^{\rm{E} 2}}_{\rho_{\rm m}}$ represent the contribution from
${j \times B}$ part and from the time
derivatives of the energy density of electro-magnetic fields, respectively.
We take the first time derivative of the magnetic fields, because this
is the leading order and the numerical treatments of the second time
derivatives are formidable. Here, we write the gravitational amplitude
as follows for later convenience,
\begin{equation}
h^{\rm TT} = h^{\rm TT}_{\rm quad} + h^{\rm TT}_{j \times B} + h^{TT}_{\rho_{\rm m}}, 
\end{equation}
where the quantities of the right hand of the equation are defined by
Eqs. (\ref{htt}), (\ref{quad}), (\ref{jB}), and (\ref{rhom}). 
It is noted that we take into account
the terms related to the artificial viscosity (see Eq. (\ref{quad}),
e.g., \cite{mm}). As mentioned, we employ an artificial viscosity of von
Neumann and Richtmyer. The concrete form of $q_{i}$ is,
\begin{equation}
q_{i} = \nabla_{i}~[l^2~\rho~(\nabla \cdot \mbox{\boldmath$v$})^2],
\end{equation}
where $i = r, \theta$ with $l$ defining the dissipation length.
In the following computations, we assume that 
observer is located in the equatorial plane since the
gravitational wave radiates most in the plane ($\alpha = \pi/2$ in
Eq. (\ref{htt})), and that the source is assumed to be located at our galactic
center ($R = 10~\rm{kpc}$).

\section{Numerical Results}

\subsection{Effects of the magnetic fields}

\begin{figure}[hbtp]
\begin{center}
\includegraphics[scale=0.64]{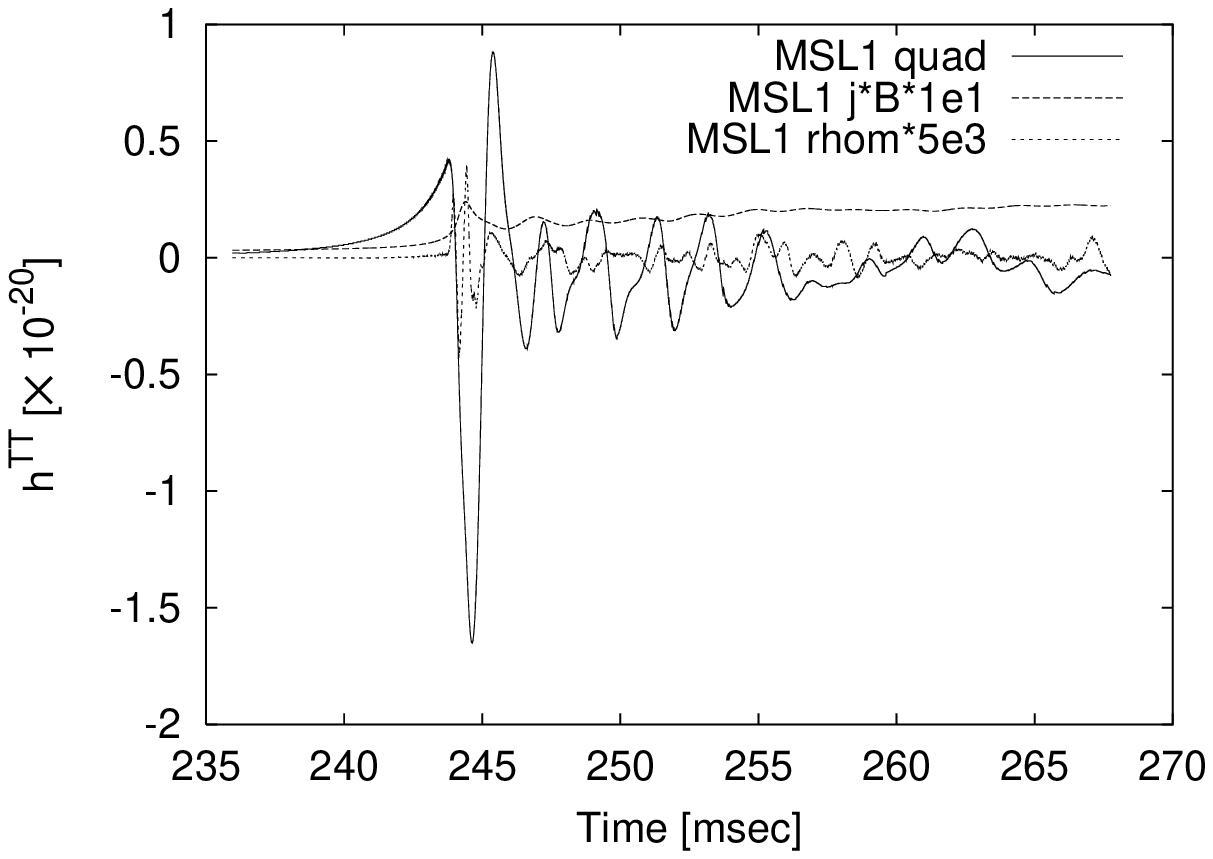}
\includegraphics[scale=0.64]{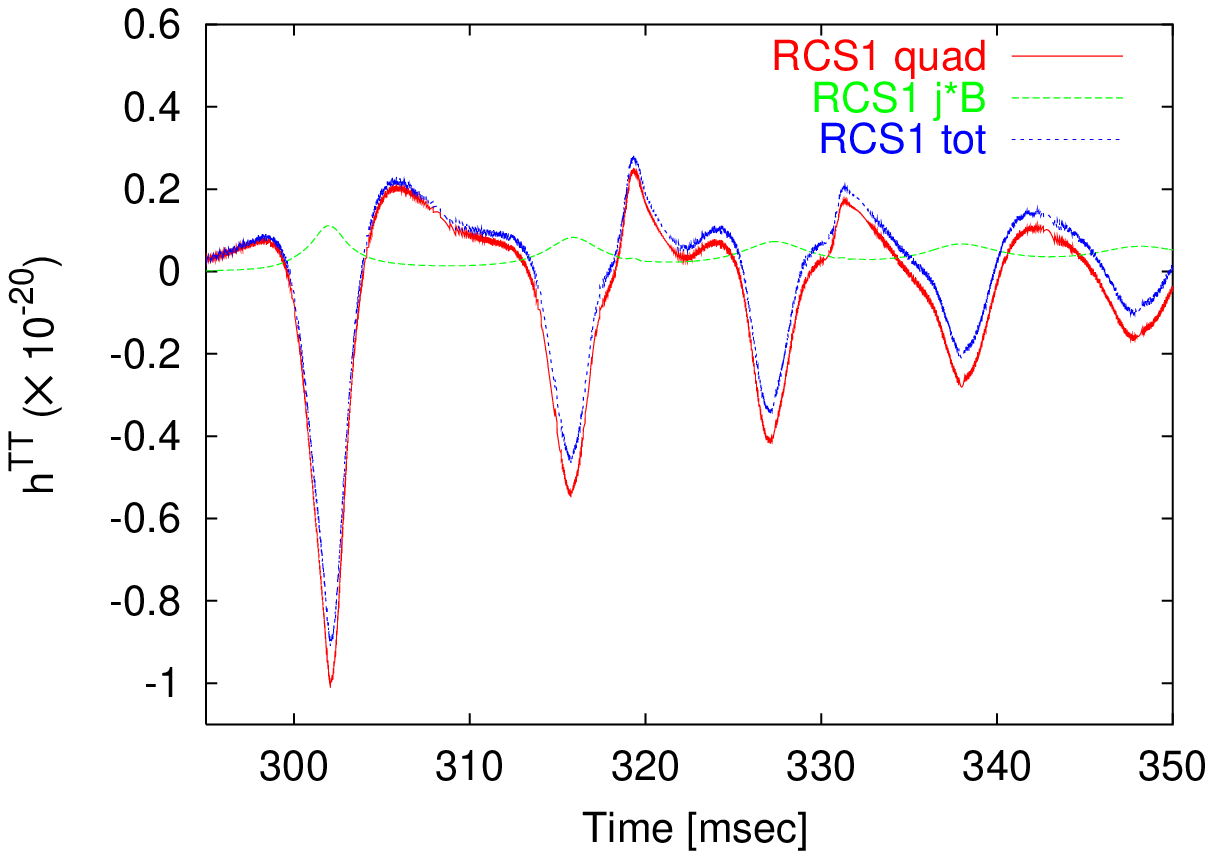}
\caption{Waveforms for
 models MSL1 (left panel), RCS1 (right panel). In the left panel,
 ``quad'', ``$j \times B$``, and ``rhom'' ($\rho_{\rm m}$) represents the
 contribution from the mass quadrupole moment, $j \times B$ part, and the
 time derivatives of the magnetic energy density ($\rho_{\rm m}$ part),
 respectively. The amplitudes are  artificially multiplied by $10^{1}$
 for $j \times B$ part and $5\times 10^{3}$ for $\rho_{\rm m}$ part to
 make their waveforms clear. From the panel, $j \times B$ part is found
 to dominate over $\rho_{\rm m}$ part. In the right panel,  ``tot''
 indicates the amplitude including the total contributions. From the
 panel, it is found that the $j \times B$ part changes at the opposite
 phase of the mass quadrupole moment. Note that the source is assumed to be located at the distance of 10 kpc. }
\label{RCS1}
\end{center}
\end{figure}

We will first show the effect of the magnetic fields on the 
amplitude of the gravitational
wave. For later convenience, the values of several important
quantities are summarized in Tables \ref{table:2} and \ref{table:3}. We
find that the amplitude is affected in the strongly magnetized models whose initial $E_{\rm m}/|W|$
is greater than 0.1 \%. It is natural because the 
amplitude contributed from the magnetic fields should be an order of $R_{\rm mag}$
in comparison with the mass quadrupole
moment component (see Eq. (\ref{modrho})), where 
\begin{equation}
R_{\rm mag} = \frac{\displaystyle{\frac{B_{\rm c}^2}{8 \pi c^2}}} {\rho_{\rm c}} \sim ~10 ~\%
~\Bigl(\frac{B_{\rm c}}{ \rm{several}~\times 10^{17}~\rm{G}}\Bigr)^2 \Bigl(\frac{\rho_{\rm c}}{10^{13}~ \rm{g}~\rm{cm}^{-3}}\Bigr)^{-1},
\label{rmag}
\end{equation}
with $B_{\rm c},\rho_{\rm c}$ being the central magnetic field and the
central density.
Thus, strongly magnetized models, whose central magnetic fields at core
bounce become as high as  
$ \sim 10^{17}$ G, can affect the amplitude.
As for the waveforms, we find that the contribution from $j \times B$
part dominates over one from the time
derivatives of the energy density of electro-magnetic fields (see the left panel of Figure \ref{RCS1}).
Furthermore, it is found that the $j \times B$ part changes at the
opposite phase of the matter quadrupole
components (see the right panel of Figure \ref{RCS1}). 
As a result, the negative parts of the amplitudes 
become less negative, while the positive parts become more positive.  
This lowers the peak amplitude at core bounce by an order
of 10 \% (see Table \ref{table:3}). If the initial strength of the magnetic
field is the same, the effect of the magnetic fields on the amplitude
becomes more prominent for the fast rotation models (compare the values of
ratio of MCS1
with RCS1, and MCS0.3 with RCS0.3 in Table \ref{table:3}). This is because 
the central density at core bounce is lowered by the faster rotation
(see Table \ref{table:2} and Eq. (\ref{rmag})).  

\begin{figure}[hbtp]
\begin{center}
\includegraphics[scale=0.64]{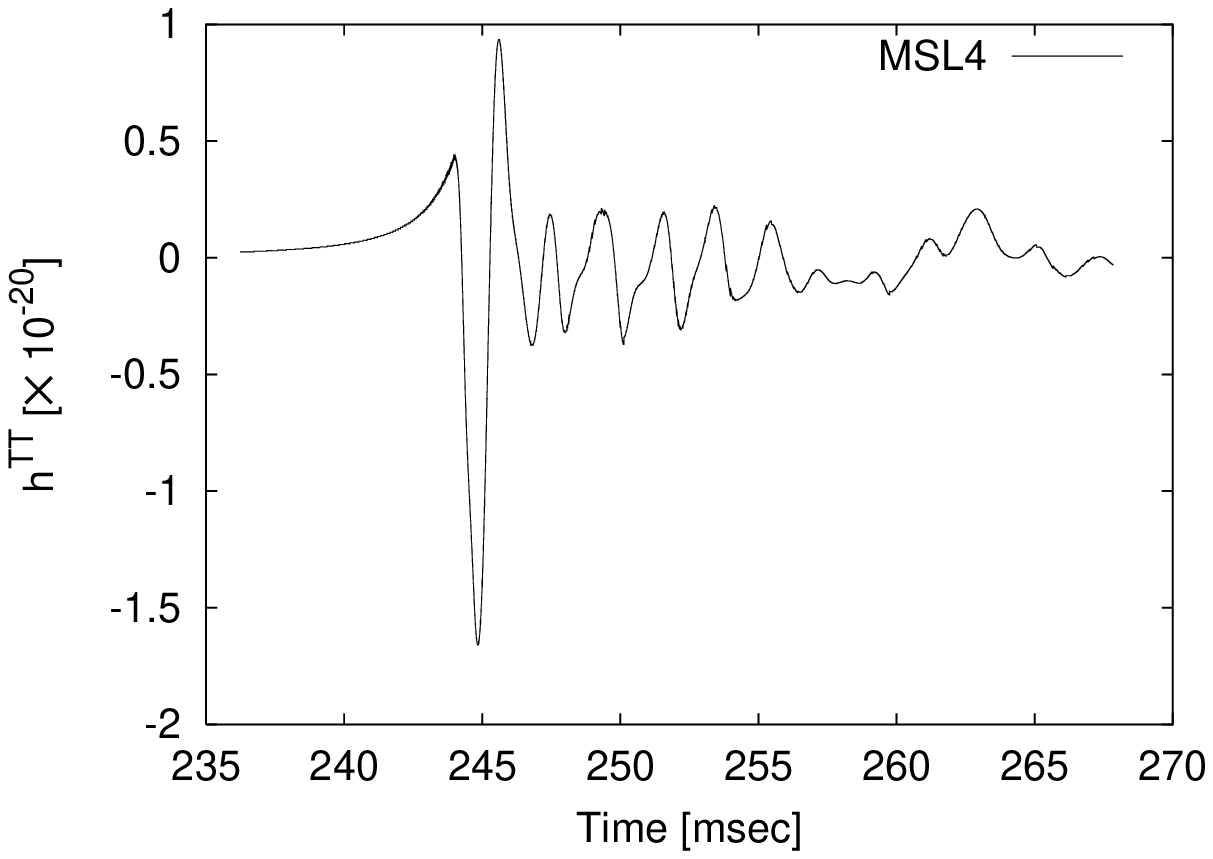}
\includegraphics[scale=0.64]{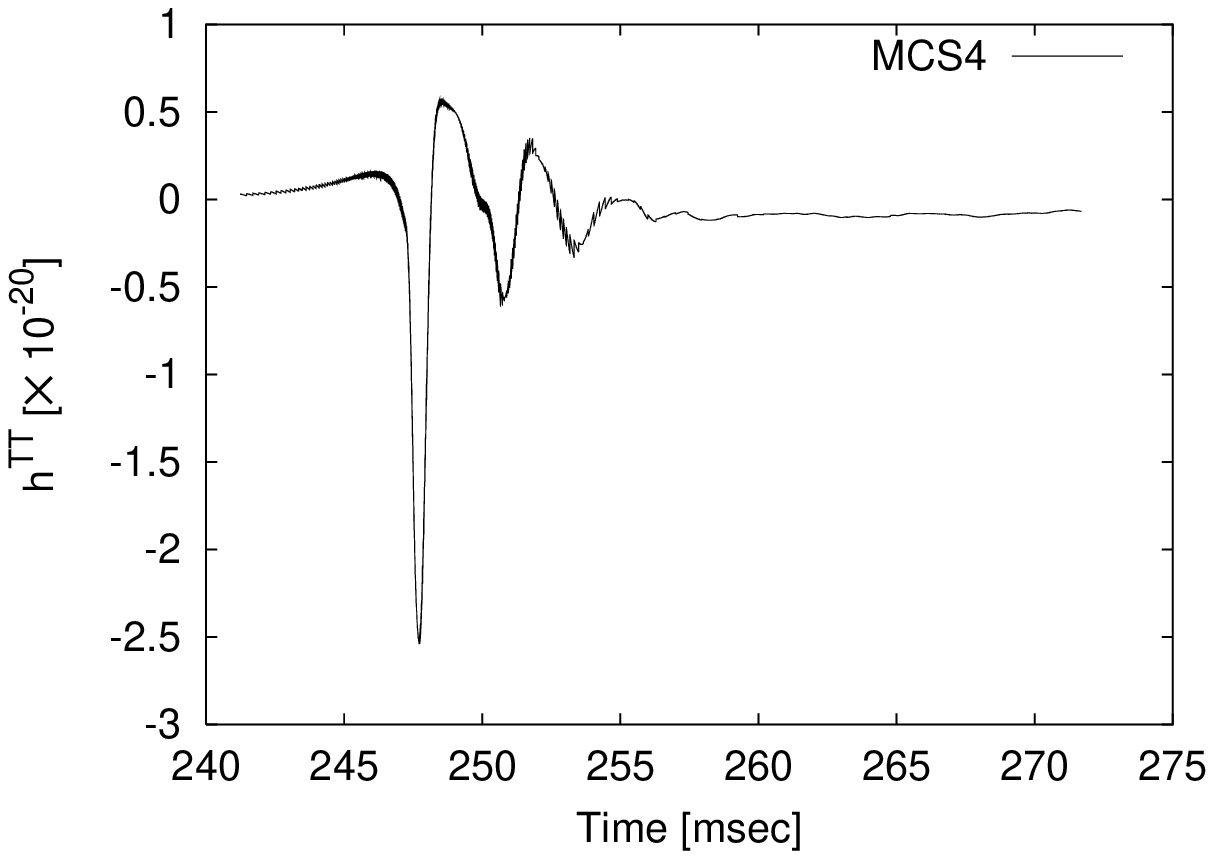}
\caption{Waveforms for
 models MSL4 (left panel), MCS4 (right panel). Note that the source is assumed to be located at the distance of 10 kpc.}
\label{typeIandII}
\end{center}
\end{figure}

For the models, whose initial $E_{\rm m}/|W|$'s are less than $0.1 \%$,
the maximum amplitude is found to be determined by
rotation as seen from Table \ref{table:2}. 
We compare the waveforms for the models MSL4 (left panel), MCS4 (right panel) in Figure
\ref{typeIandII}. If we follow the commonly used categories
 of the waveforms \cite{mm,zweg}, the left panel shows type I behavior 
(relatively
 shorter intervals of the spikes) and the right
 panel type II behavior (longer intervals of the spikes). 
What determines the difference is the initial rotation rate and degree of differential rotation. 
It is found that type II occurs for the models with the strong differential
rotation with a cylindrical rotation law as in the pure rotation case 
by Kotake {\it et al}. \cite{kotakegw} and 
Ott {\it et al}. \cite{ott}. Furthermore, the
sign of the values of the second peaks is found to be negative for the strongly
differentially rotating models with a cylindrical rotation law, and positive
for the other models (see Table \ref{table:4}).  Note that by the
``second peak'' we mean where the absolute amplitude is second largest. 
Due to the $j \times B$ part, the second peaks with the positive values
 become more positive, while those with the negative values become
 generally less negative (see Table \ref{table:4}).

\subsection{Effects of the equation of states}
We compare the models MSL4 with MSL4-LS in this section. We repeat that 
these models differ from its employed EOS's. Other than this, any
conditions, such as rotation, magnetic field, and the employed
microphysics, are the same.  
Therefore the differences in the following purely reflect the influences of
the employed EOS's.
The most important difference of the two EOS's is the
stiffness of the EOS's at nuclear matter density. 
In fact, the value of the incompressibility of the relativistic EOS (K =
281 MeV) is larger than that of 
Lattimer-Swesty EOS (K = 180 MeV). This means that the Lattimer-Swesty EOS is softer
than the relativistic EOS.
We note that there are three choices with different
values of (K = 180, 220, 375 MeV) in the Lattimer-Swesty EOS. We take the
most softest one in this calculation. This is because the value has been often employed for core collapse supernova studies.

\begin{figure}[hbtp]
\begin{center}
\includegraphics[scale=0.64]{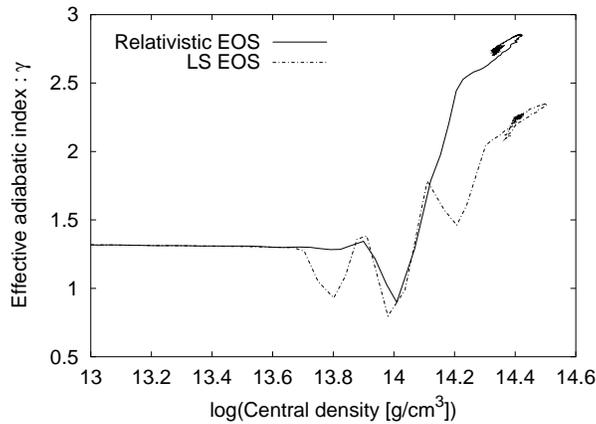}
\caption{Relation between the central density and the effective
 adiabatic index $\gamma$ at near core bounce. Relativistic EOS, LS EOS in the figure
 represents the relation taken from the values from the models MSL4 and
 MSL4-LS, respectively. The swirls at the density of $\sim 10^{14.4}~\rm{g}~\rm{cm}^{-3}$ are due to the reexpansion of the core after core bounce.} 
\label{gamma_hikaku}
\end{center}
\end{figure}

A core bounce occurs when the central density reaches its peak of 
$3.2 \times 10^{14}~\rm{g}~{\rm cm}^{-3}$, $2.6 \times
10^{14}~\rm{g}~{\rm cm}^{-3}$ at $t_{\rm b} = $ 215~ms, 243~ms for the models 
MSL4-LS, MSL4, respectively.
The earlier core bounce with the higher central density  
for the model MSL4-LS is due to the softness of the EOS. The softness of the
Lattimer-Swesty EOS can be seen from Figure \ref{gamma_hikaku}.
The soft EOS results in the more smaller inner core of $M_{\rm ic} =
0.69 M_{\odot}$ than that of $M_{\rm ic} = 0.83 M_{\odot}$ by the
relativistic EOS (see Table \ref{table:2}). This result can be understood as follows. The mass of
inner core at core bounce is proportional to the square of the lepton fraction,
$Y_{l}$. The electron capture with the
neutrino emission proceeds further for the
soft EOS because the core can contract more compact.
As a result, the lepton fraction at core bounce becomes smaller, which
results in the smaller mass of the inner core.

\begin{figure}[hbtp]
\begin{center}
\includegraphics[scale=0.64]{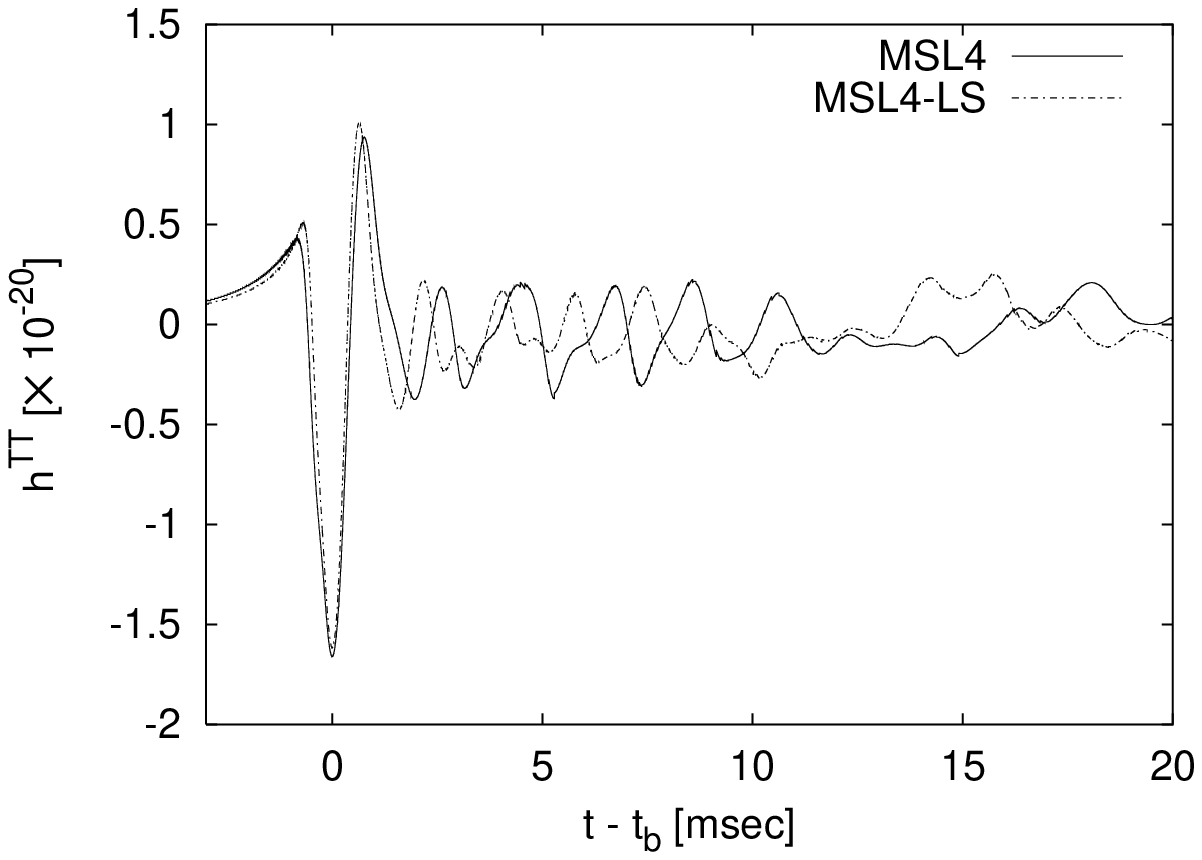}
\includegraphics[scale=0.64]{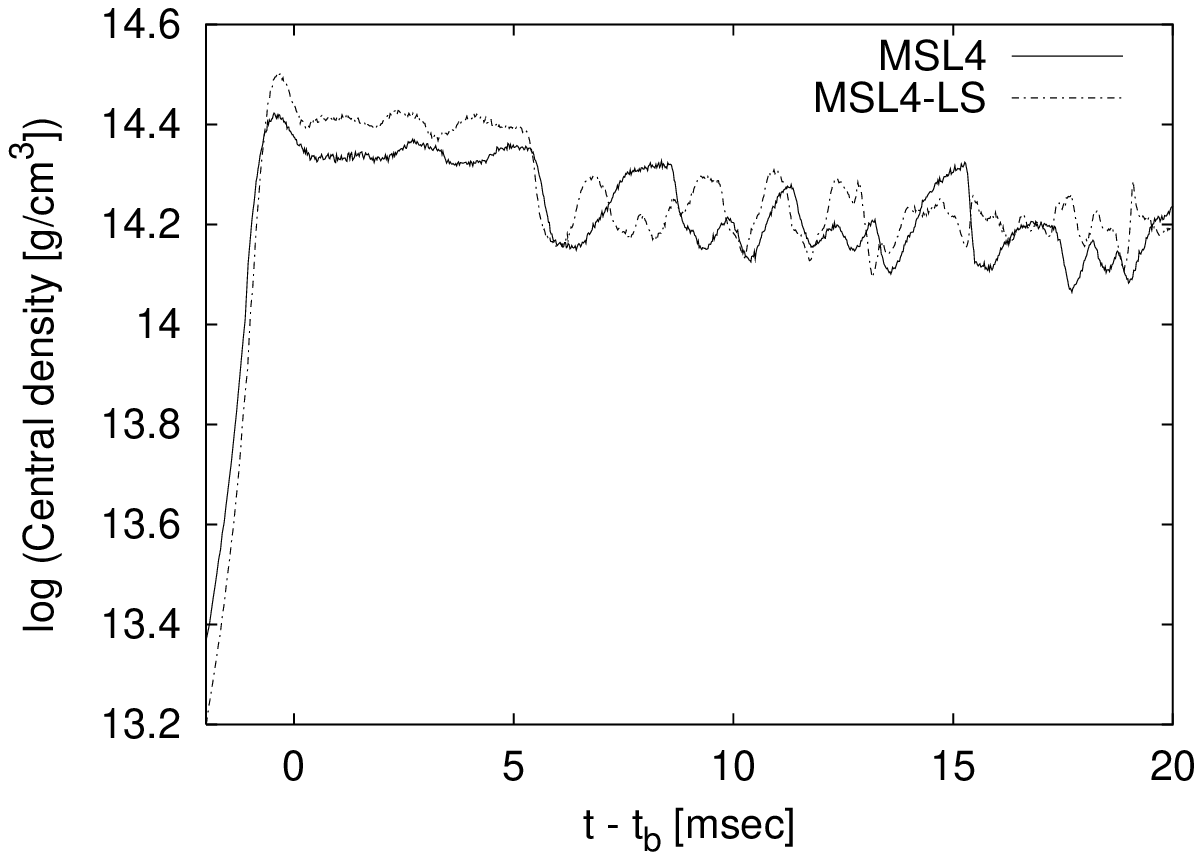}
\caption{Waveforms (left
 panel) and the time evolution of the 
 central density (right panel) for
 models MSL4 (left panel), MSL4-LS (right panel). All times are relative
 to the time of bounce ($t_{\rm b}$) in this figure. Note that the source is assumed to be located at the distance of 10 kpc.}
\label{eoshikaku}
\end{center}
\end{figure}

In the left panel of Figure \ref{eoshikaku}, the gravitational waveforms for the models are
presented. Shorter burst intervals for the model MSL4-LS are also 
due to the softness of the EOS. The duration of the
burst are related to the typical dynamical timescale $\tau_{\rm dyn} \sim
1/ \sqrt{G \rho_{\rm c}}$, where $\rho_{\rm c}$ is the central
density. Since the 
values of $\rho_{\rm c}$ at the core bounce and the subsequent
oscillations are larger for the soft EOS (see the right panel of Figure \ref{eoshikaku}), the intervals of the burst
become short. Although the collapse dynamics is changed by the differences of
the EOS's, it is found that the values of 
the maximum amplitudes for the two models do not differ significantly
(see Table \ref{table:2}). This can be understood as follows. 
The amplitude of a gravitational wave is
roughly proportional to the inverse square of the typical dynamical time scale,
(see Eq. (\ref{quadru})). Since $t_{\rm dyn}$ is proportional to the inverse
root of the central density, the amplitude becomes larger for the soft
EOS by this factor. On the other hand, the amplitude is proportional
to the value of the quadrupole moment, which becomes in turn small for
the soft EOS due to the smaller inner core. The maximum amplitude is determined by the competition between
these factors. As a result, the amplitude becomes almost the same for
the Lattimer-Swesty EOS and the relativistic EOS.     

In our calculations, we note that 
there are no models that correspond to the so-called
type III waveform, which occurs only when a EOS is very soft before
core bounce \cite{zweg}.
This is because the relativistic EOS and Lattimer-Swesty EOS is not so
soft in the corresponding regime.

\subsection{The gravitational wave properties in the weakly
  magnetized and slowly rotating model}
\begin{figure}[hbt]
\begin{center}
\includegraphics[scale=0.64]{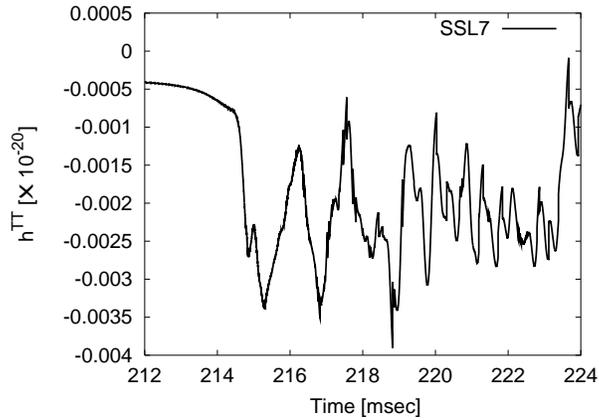}
%\vspace{0.5 cm}
\caption{Waveform for model SSL7, which is based on the recent stellar
 evolution calculation. The amplitudes are about three orders of
 magnitudes smaller than the other models. Note that the source is assumed to be located at the distance of 10 kpc.}
\label{SSL7}
\end{center}
\end{figure}

We show the properties of the waveform with collapse
dynamics in the model SSL7. It is noted that the initial
condition for the model is based on the recent stellar evolution
calculation \cite{heger03}.  
We repeat that the initial $T/|W|$ and $E_{\rm m}/|W|$ are much smaller
than the other models.
The model bounces at $t = 215$ msec at a central density of $3.0 \times
10^{14} \rm{g}~\rm{cm}^{-3}$. It occurs not by rotation or magnetic
fields but by the stiffening of the equation of state. 
After the core bounce, very weak
signals are lasting (see Fig. \ref{SSL7}). The gravitational 
amplitudes from the bounce and later reexpansion phases are three orders
of magnitude smaller than the other models. 
The timescale of the spikes ($\sim$ msec) is determined by the small
scale non-radial motions behind the shock wave. 

%This result is consistent with the recent study by Ott {\it et al.} \cite{ott}%.

\subsection{Maximum amplitude and second peak}   
\begin{figure}
\includegraphics{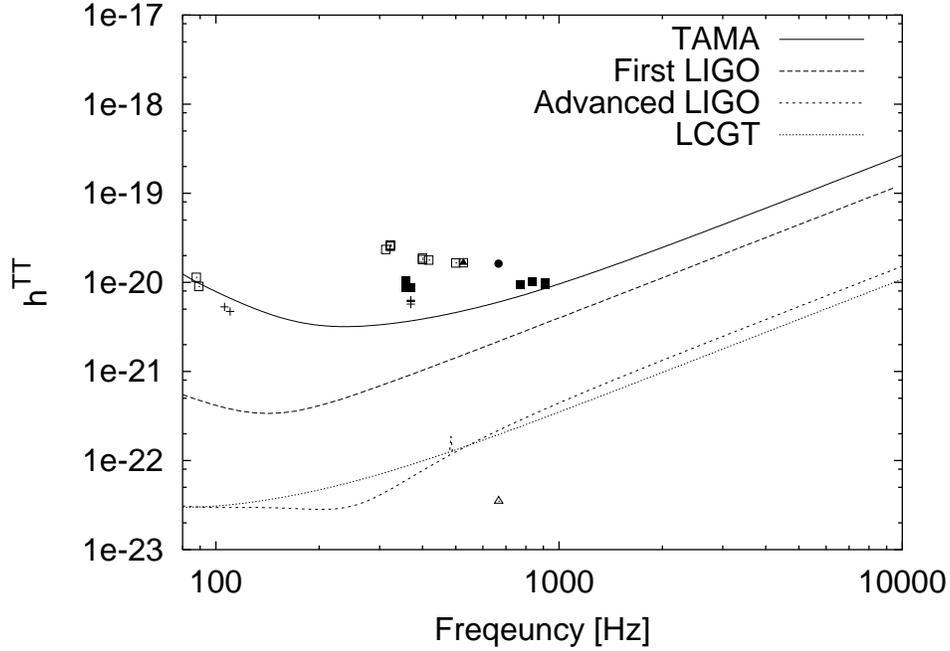}
\caption{Detection limits of TAMA \cite{tama}, first LIGO
 \cite{firstligo}, advanced LIGO \cite{advancedligo}, and Large-scale
 Cryogenic Gravitational wave Telescope (LCGT) \cite{lcgt} with
 the expected amplitudes from numerical simulations. The open squares
 represent the maximum amplitudes for all the models, except for models MSL4
 (closed triangle) and MSL4-LS (closed circle). On the other hand,
 the pluses and the closed squares represent the amplitudes of second
 peaks for models with strong differential rotation with a cylindrical
 rotation law and for the other models, respectively. Open triangle
 represents the maximum amplitude for model SSL7. We estimate the characteristic
 frequencies by the inverse of duration periods of the corresponding 
 peaks. 
Note that the source is assumed
 to be located at the distance of 10 kpc.}
\label{detect}
\end{figure}

In Figure \ref{detect}, the absolute values of the maximum amplitudes and
second peaks are presented. Note that the data points are confined to
small regions although we explore the wide range of the initial magnetic
field strength. This is because the magnetic fields lower the peak
ampitudes by $\sim 10 \%$ but do not change the typical frequencies of
the gravitational waves significantly. Thus the maximum amplitudes and
the typical frequencies are mainly determined by rotation. The maximum
amplitudes are mainly clustered in the two regions (see the open squares
in Figure \ref{detect}). One is around $\sim 100$ Hz for the models whose names
begin with ``R'' (rapidly rotating models) and another is around from $\sim 300$ Hz to $\sim 500$ Hz
for the models whose names begin with ``M'' (moderately rotating models). The values of the maximum amplitudes are in
the range of $ 3.5 \times 10^{-23} \leq h^{\rm TT} \leq 2.6 \times
10^{-20} $. The smallest value is from model SSL7, whose
initial model is taken from the recent stellar evolution
calculation. If the model is correct, 
it should be difficult to detect the gravitational wave even for the next
generation detectors such as advanced LIGO and LCGT unless a supernova
occurs very close to us (see Figure \ref{detect}). 
When we compare the maximum amplitudes from 
model MSL4 (relativistic EOS) with that from model MSL4-LS (Lattimer-Swesty
EOS), no significant differences are found, except that the
typical frequency for LS EOS becomes slightly higher due to the softness
of the EOS
(see Figure \ref{detect}). 
Therefore the features of the gravitational wave are almost 
independent of the realistic EOS's. 
 
As pointed out earlier, we found that the signs of the values of the
second peaks are negative for models with strongly differential rotation 
with a cylindrical rotation law and positive for the others (see Table
\ref{table:3}). 
The absolute amplitudes of the second peak are also presented in Figure
\ref{detect}. As shown, they are within the detection limit of first LIGO for
a source at a distance of 10 kpc, although the absolute values of the 
negative values are reduced by the magnetic fields. 
In addition, it is quite likely that the detectors in the next
generation, such as advanced LIGO and LCGT, to detect the difference.
We will obtain information about the angular momentum distribution of
evolved massive stars, if we can detect the differences by observations
of the gravitational wave. However, it seems difficult to get
information about the distributions of the magnetic fields only by the
gravitational wave from the magnetorotational core collapse. We note that the effects of the magnetic fields on
the signals may become important after the formation of the strongly
magnetized star. As pointed \cite{ioka,miralles}, the gravitational waves, which
are detectable for interferometers such as LIGO, may be emitted by the
global rearrangement of the strongly magnetic fields ($B > 10^{16} G$)
in the core.       
%%%%%%%%%%%%%%%%%%%%%%%%%%%%%%%%%%%%%%%%%%%%%%%%%%%%%%%%%%%%%%%%
\begin{table}
\caption{Summary of important quantities for all models. $t_{\rm b}$ is
 the time of bounce, $\rho_{\rm{max b}}$ is the maximum density at
 bounce, $M_{\rm i.c \,\, b} $ is the mass of inner core at bounce,
 $T/|W|_{\rm final}, E_{\rm m}/|W|_{\rm final}$ is the final ratio of
 rotational and magnetic energies to
 gravitational energy of the core, respectively.
 $\Delta t$ is the duration time full width at half maximum of the first burst, and
 $|h^{\rm{TT}}|_{\rm max}$ is the maximum amplitude, including
 the contributions from the electromagnetic fields.  Note that we speak of the inner core, where the matter falls
 subsonically, which corresponds to the unshocked region after core
 bounce. } 
\label{table:2}
\begin{center}
\begin{tabular}{cccccccc} \hline \hline
Model & $t_{\rm b}$ & $\rho_{\rm{max b}}$  & $M_{\rm i.c \, b}$ 
 & $T/|W|_{\rm final} $ & $E_{\rm m}/|W|_{\rm final}$ 
& $\Delta t$   & $|h^{\rm{TT}}|_{\rm max}$ \\
      & (ms)   & ($10^{14}\,\,\rm{g}\,\,\rm{cm}^{-3}$) & $(M_{\odot})$ &
 $(\% )$& $(\%)$ & (ms)
 & $(10^{-20})$ \\ \hline
 SSL7 & 214.6  & 2.97  & 0.61  & $1.0 \times 10^{-2}$   & $3.5 \times 10^{-8}$   &  0.75  &  $3.5 \times 10^{-3}$     \\ 
 MSL4 & 242.8  & 2.61  & 0.83  & 8.7 & $4.3 \times 10^{-5}$  & 0.74   &  1.66     \\ 
 MSL2 & 242.9  & 2.66  & 0.83  & 8.6 & $4.2 \times 10^{-3}$  & 0.73   &  1.66\\ 
 MSL1 & 242.7  & 2.65  & 0.83  & 8.6 & $3.9 \times 10^{-2}$  & 0.73   &  1.65 \\    
 MSS4 & 241.8  & 1.72  & 0.87  & 9.0 & $2.9 \times 10^{-4}$  & 0.58   &  1.78 \\
 MSS2 & 242.0  & 1.71  & 0.87  & 9.0 & $2.9 \times 10^{-2}$  & 0.58   &  1.80 \\
 MSS1 & 242.1  & 2.10  & 0.87  & 9.1 & $2.4 \times 10^{-1}$  & 0.58   & 1.87\\  
 MCS4 & 245.1  & 1.45  & 0.91  & 9.9 & $3.0 \times 10^{-4}$  & 0.53   & 2.54 \\
 MCS2 & 244.1  & 1.55  & 0.91  &10.0 & $2.7 \times 10^{-2}$  & 0.53   & 2.58 \\
 MCS1 & 244.2  & 1.88  & 0.91  &10.1 & $1.8 \times 10^{-1}$  & 0.53   & 2.63 \\
 MCS0.3 &245.3 & 2.25  & 0.91  &10.0 & $5.8 \times 10^{-1}$  & 0.53   & 2.34 \\
 RCS1 & 302.2  & 0.17  & 1.11  &12.2 & $3.1 \times 10^{-1}$  & 2.03   & 0.90                                    \\ 
 RCS0.3& 320.5 & 0.84  & 1.14  &12.2 & $8.0 \times 10^{-1}$  & 1.76   & 1.14 
 \\ \hline
 MSL4-LS & 215.4 & 3.20 & 0.69  & 8.2    & $3.4 \times 10^{-5}$                      & 0.63   & 1.62
\\ \hline \hline
\end{tabular}
\end{center}
\end{table}

\begin{table}
\caption{Effects of the strong magnetic fields on the maximum
 amplitudes. Note that all the values of the amplitudes of the
 gravitational wave are given in unit of $10^{-20}$ in this table. }
\label{table:3}
\begin{center} 
\begin{tabular}{lccccrccc} \hline \hline
Model & $h^{\rm TT}_{\rm quad} $  & $h^{\rm TT}_{ j \times B} $   
 & $h^{\rm TT}_{\rho_{\rm m}} $  & $ |h^{\rm TT}_{ j \times B}/ h^{\rm TT}_{\rm quad}| (\%)$      
 \\ \hline
 MSL1   & - 1.65   & $2.40 \times 10^{-2}$  & $ - 8.56 \times 10^{-4}$&
 1.5    \\
 MSS1   & - 2.16   & $2.95 \times 10^{-1}$  &  $- 1.04 \times 10^{-3}$&
 13.8   \\
 MCS1   & - 2.69   & $6.46 \times 10^{-2}$  &  $- 1.01 \times 10^{-3}$& 
 2.4    \\
 MCS0.3 & - 2.55   & $2.13 \times 10^{-1}$  &  $- 4.63 \times 10^{-3}$&
 8.3    \\
 RCS1   & - 1.01   & $1.11 \times 10^{-1}$  &  $- 2.45 \times 10^{-3}$&
 10.9   \\
 RCS0.3 & - 1.53   & $3.83 \times 10^{-1}$  &  $- 1.92 \times 10^{-3} $&       
 25.0   
  \\\hline \hline
\end{tabular}
\end{center}
\end{table}
 \begin{table}
\caption{Some characteristic quantities for the waveform analysis. The names of
 the models whose initial rotation law is cylindrical with strong
 differential rotation are written in bold letters.
$T^{I}_{\rm osc}$ and $T^{II}_{\rm osc}$ the first and second
 oscillation period of the inner core, respectively.  $h^{TT}_{\rm second}$
 is the amplitude of gravitational wave at the second peak.}
\label{table:4}
\begin{center}
\begin{tabular}{cccc} \hline \hline
Model & $T^{I}_{\rm osc}$   & $T^{II}_{\rm osc}$&  $h^{TT}_{\rm second}$ \\
     & [ms]       & [ms]   & $[10^{-20}]$ \\\hline
 MSL4 & 1.9       & 1.3        & 0.93 \\ 
 MSL2 & 1.9       & 1.1        & 0.93   \\              
 MSL1 & 2.0       & 1.2        & 1.02   \\
 MSS4 & 2.4       & 2.8        & 0.87  \\
 MSS2 & 2.5       & 2.7           & 0.88     \\  
 MSS1 & 2.5       & 2.8            & 1.03 \\
{\bf MCS4} & 3.1         & 2.7    & - 0.62  \\
 {\bf MCS2} & 3.1        & 2.7          & - 0.61    \\ 
 {\bf MCS1} & 3.1        & 2.7        &    - 0.63  \\
 {\bf MCS0.3}  & 3.2     & 2.7         & - 0.57           \\
 {\bf RCS1 }    & 11.2     & 9.1      & - 0.47            \\
 {\bf RCS0.3}  &  11.4     & 9.2          & - 0.53         
\\ \hline
 MSL4-LS & 1.5   & 1.1         & 1.00  
\\\hline \hline
\end{tabular}
\end{center}
\end{table} 
\section{Conclusion}
We have done a series of two-dimensional magnetohydrodynamic simulations
of the rotational collapse of a supernova core and calculated
gravitational waveforms. We have modified the conventional quadrupole
formula in order to include the contributions from the electromagnetic
fields. Recent stellar evolution calculations imply that the magnetic
field of the toroidal component is much stronger than that of the
 poloidal ones at the presupernova stage. In this study, we 
 systematically investigate the effects of the toroidal magnetic fields
 on the gravitational waveforms.
We have employed the two kinds of the realistic equation of
states, which are often used in the supernova simulations. By so doing,
we have investigated the effects of the equation of sates on the
gravitational signals. As for the microphysics, 
we took into account
electron capture and neutrino transport in an approximate method.
We found the following.\\  

(1). 
Effects of the magnetic field on the amplitude and the waveform of the
gravitational waves appear for the strongly
magnetized models, whose initial $E_{\rm m}/|W|$'s are greater than $0.1
\%$.  As for the contributions from the magnetic field, the $j \times B$
part is found to dominate over the time derivatives of the 
magnetic energy. Since $j \times B$ part changes with the opposite phase
in comparison with the mass quadrupole moment components, the maximum amplitudes are lowered by an order of $10
\%$ for the strongly magnetized models. 
However, the maximum amplitudes are
mostly within the detection limits of the detectors of TAMA and first
LIGO if a source is located at a distance
of 10 kpc.

(2) The maximum amplitudes do not change significantly if we employ the
two realistic equation of states commonly used in the supernova
simulations. Since Lattimer-Swesty EOS is softer than the relativistic
EOS, the mass of the inner core for the Lattimer-Swesty EOS becomes small, which
reduces the mass quadrupole moments at core bounce. 
On the other hand, the softness allows the core to contract deeply,
which makes the central density larger at core bounce. By the competition of
these factors, the maximum amplitude 
remains almost the same between the two realistic EOS's. We note the
difference that the typical frequencies of the gravitational waves for the
Lattimer-Swesty EOS become higher due to the softness of the
EOS. These effects of the EOS's on the gravitational waves are found to
be independent of those of the magnetic fields, since the strong magnetic fields change the maximum amplitudes and do not change the typical frequencies of the gravitational waves. 
 
(3) For the weakly magnetized models, whose initial $E_{\rm m}/|W|$'s
are smaller than $0.1 \%$, the gravitational wave amplitude and the
waveform are determined by rotation as in
the pure rotation case \cite{kotakegw}. The waveforms are categorized into the conventional criteria
\cite{zweg}.
The maximum amplitudes are within the detection limits for
the detectors of TAMA and first LIGO for a supernova at a distance of 10
kpc.
From the model based on the recent stellar evolution calculation, it seems difficult to detect the gravitational wave unless a supernova occurs very close to
us. Type III does not occur if the two realistic EOS's are employed.

(4) The signs of the values of the second peaks are found to be negative for the strongly
differentially rotating models with a cylindrical rotation law, and positive
for the other models.  
Due to the $j \times B$ part of the electromagnetic contributions, the positive values of the second peak
become more positive, while the negative values become less negative. However the absolute values of the second peaks are within the
detection limit of the first LIGO for a supernova at a distance of 10 kpc. It means that it will give us
information about the angular momentum distribution of massive evolved
star if a supernova occurs at our galactic center. On the other hand, it
seems difficult to extract the information about the magnetic field only
by the gravitational wave from a magnetorotational collapse of a
supernova. By the lack of the significant features by the magnetic fields in the gravitational signals in addition to those by rotation, a weak upper limit of the initial magnetic field strength of $ \sim 10^{14}$ G in the central core prior to gravitational collapse may be put.

Finally, we state some discussions based on the results in this
study. In all the models, the final rotation rates are within the critical
value where Maclourion spheroids become secularly unstable against
triaxial perturbations ($T/|W|_{\rm final} <$ $13.75 $\%). Therefore,
the axial symmetry assumed in this study may be justified. It is also
noted that 
Rampp et al. \cite{rampp}, who performed the three-dimensional (3D)
rotational core-collapse simulations, suggest that the features of the
gravitational signals do not differ significantly from those by the
two-dimensional simulations. However, it is necessary to perform the 3D
simulations for the more realistic estimation of the gravitational wave signals.  
As for the magnetic field contributions, the entanglement of the field
lines in the azimuthal direction may suppress the gravitational
radiation for a given rotation rate. Furthermore, it is noted that the
non-axisymmetric motions of the matter produce the cross modes of the
gravitational waves, which vanish in case of axial symmetry \cite{mujan,fryer03}. From the recent results by Fryer \&
Warren (2003) \cite{fryer03}, who did 3D rotational core-collapse
simulations, the maximum amplitude of the cross modes was presented to be $\sim 10^{-22}$
at several $100$ Hz for a supernova in our galactic center. Although this value is
smaller than that of the plus modes, the value seems to be within the
detection limits of the gravitational wave detectors in the next generation,
such as advanced LIGO and LCGT. If the characteristics of the waveforms
between the plus and cross modes are detected, we think that it may be a
good tool to extract information not only about rotation but also about
magnetic fields in the evolved massive star. In order to investigate this, we are currently preparing for the 3D MHD simulations
(H. Sawai et al. 2004 in preparation).

Recent observations imply that gamma-ray bursts (GRBs) are associated
with core-collapse supernovae \cite{galama,stanek}. Although the failed supernova or the
so-called collapsar model are supposed to be probable \cite{collapsar1,collapsar2}, it is still
controversial what powers GRBs. As the energy source for fireballs, the energy deposition by neutrinos emitted from the accretion disk or by the MHD process which extracts the angular momentum of the rotating black hole are considered to be important. 
MacFadyen {\it et al}. \cite{collapsar3} reported that   
matter with the mass of $\sim 0.1 - 5 M_{\odot}$ falls into the central
black hole during minutes to hours, which suggests the duration for the
activity of GRBs. It is noted that no magnetic fields are considered
in their calculations. When we boldly extrapolate the central magnetic
fields in our strongest magnetized models, which become as high as $\sim
10^{17}$ G in the central region, to the model of GRBs, it may not seem so
unnatural to imagine the infall of the strongly magnetized material into 
the black hole. If the magnetized matter falls into the black hole
anisotropically, we expect the emission of the gravitational waves. In order to calculate the amplitudes, the formula, which we derived in this paper, will be useful. 
Here it should be noted that the fully general relativistic
magnetohydrodynamic simulations with the multi-dimensional neutrino
transport are required for the reliable estimation to determine the
masses of the black hole and the accretion disk.
Therefore further advancements in the numerical simulations are necessary for the quantitative discussions.

Recently, one group claims the detection of linear polarization of $\sim
80$ \% in a GRB \cite{coburn}, however, the
other group claims that the polarization is less than $\sim 4$ \% \cite{rutledge}. The discrepancy may come from the difference in the way of data analysis. If the former is correct, it may seem consistent with our results, which predict rather coherent magnetic fields in the vicinity of the central core.   

  As stated earlier,  M\"{u}ller {\it {et al.}} \cite{muller03} recently reported that the
  maximum amplitude is not obtained at the core bounce but at  
  the later phases when the anisotropic neutrino radiation 
  becomes active (see, also, \cite{burohey,mujan}).
  If the degree of the anisotropy of neutrino radiation 
  could be observed from the
  gravitational radiation, it will help to understand the explosion
  mechanism itself. This is because anisotropic
  neutrino  radiation induced by rotation is likely to produce an
  asymmetric explosion \cite{shimi01,kotakeaniso,fryer03} 
  as suggested by observation of SN 1987 A
  \cite{wang96,pun,wang01,wang02}. Unfortunately, this is beyond our scope now to estimate the effects of anisotropic neutrino radiation on the gravitational signals by our crude treatment of the neutrino transport. For the purpose, we are now developing a two-dimensional neutrino transport code, which
  is indispensable for 
  the estimation of the gravitational wave from the anisotropic
  neutrino radiation.

\section*{Acknowledgments}
We are grateful to K. Numata and M. Ando for helpful discussions.
K.K would like to be thankful to M. Shimizu and M. Oguri for supporting
computer environments. 
The numerical calculations were partially done on the
supercomputers in RIKEN and KEK (KEK supercomputer Projects No.02-87 and
No.03-92). This work was partially supported by 
Grants-in-Aid for the Scientific Research from the Ministry of
Education, Science and Culture of Japan through No.S 14102004, No.
14079202, and No. 14740166.
      
\appendix


\clearpage


\end{document}